# Robust and Scalable Entity Alignment in Big Data

James Flamino[†]
Systems & Technology Research
flamij@rpi.edu

Christopher Abriola
Systems & Technology Research
christopher.abriola@stresearch.com

Ben Zimmerman
Systems & Technology Research
benjamin.zimmerman@stresearch.com

Zhongheng Li
Systems & Technology Research
zhonghen.li@stresearch.com

Joel Douglas
Systems & Technology Research
joel.douglas@stresearch.com

## ABSTRACT

Entity alignment has always had significant uses within a multitude of diverse scientific fields. In particular, the concept of matching entities across networks has grown in significance in the world of social science as communicative networks such as social media have expanded in scale and popularity. With the advent of big data, there is a growing need to provide analysis on graphs of massive scale. However, with millions of nodes and billions of edges, the idea of alignment between a myriad of graphs of similar scale using features extracted from potentially sparse or incomplete datasets becomes daunting.

In this paper we will propose a solution to the issue of large-scale alignments in the form of a multi-step pipeline. Within this pipeline we introduce scalable feature extraction for robust temporal attributes, accompanied by novel and efficient clustering algorithms in order to find groupings of similar nodes across graphs. The features and their clusters are fed into a versatile alignment stage that accurately identifies partner nodes among millions of possible matches. Our results show that the pipeline can process large data sets, achieving efficient runtimes within the memory constraints.

**ACM Reference format:**
James Flamino, Christopher Abriola, Ben Zimmerman, Zhonghen Li and Joel Douglas. 2020. Robust and Scalable Entity Alignment in Big Data. *9 pages.*

## 1 Introduction

Entity alignment refers to the process of identifying an entity and matching that same entity across disparate data streams. The application of entity alignment has seen significant use in a myriad of fields, including data management, data mining, biomedicine, and machine learning [15, 19]. A particularly common and compelling application of entity alignment can be found in social media analysis, where matching users across different social media platforms by correlating account activity can be very useful for research on cross-platform information propagation [9, 12, 16]. Entity alignment is also quite useful within the field of the natural language processing, where mapping words and semantic meaning between languages can provide invaluable labeling for machine translation services [3]. And while work in the field of entity alignment is comprehensive, the advent of big data has made the need for these kind of applications for exponentially growing event streams even more apparent.

Our work focuses on cases where alignment relies on correlating account activity, also referred to as "events" based on features extracted from these activities. With tens of millions of unique entities and billions of events for target datasets, the alignment between massive event streams quickly becomes challenging. As the sizes of these datasets grow, entity ontology diversifies and expands as well, making node-to-node feature correlation increasingly complex and expensive. Ultimately, entity

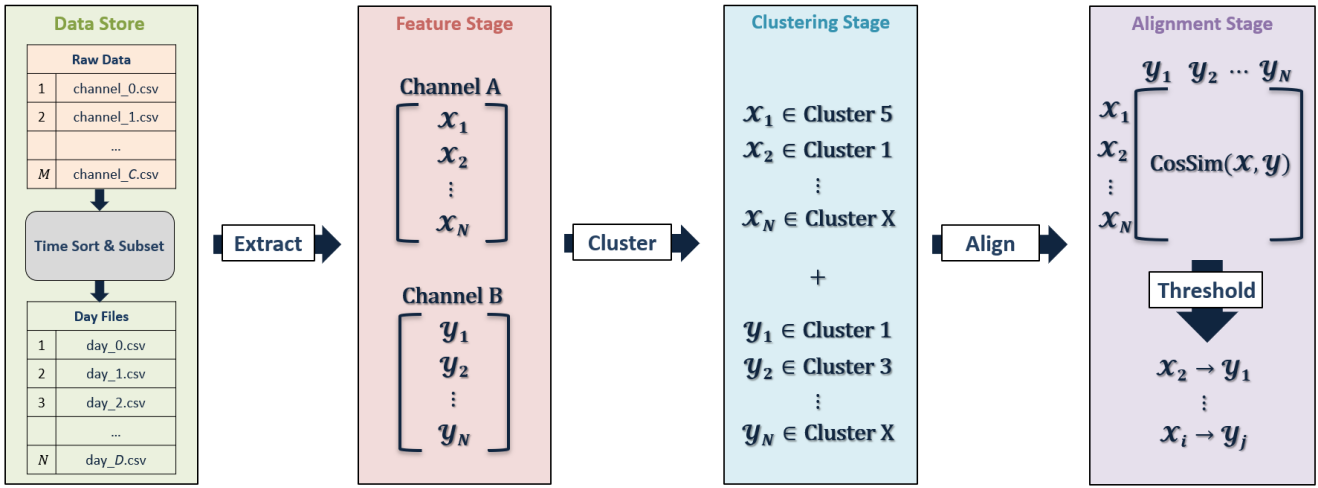

**Figure 1: Full Entity Alignment Pipeline**

alignment faces the difficult problem of matching unique nodes across ever-growing datasets.

In this paper, we study the problem of entity alignment within the scope of this big data problem and propose a solution in the form of a scalable, multi-step pipeline. In Section 4, we begin our description of this pipeline by discussing the paradigm of efficient feature extraction of common (and often universal) attributes in data streams from different graphs, such as temporal features. In Section 5 we outline an accompanying clustering architecture, which implements customizable methods of efficient clustering oriented toward grouping similar entities in order to greatly reduce the number of comparisons that need to be made between data sources. In Section 6 we examine methods for alignment given our types of features and clusters and introduce enhancements through the use of alignment likelihood scores. We evaluate our pipeline using a set of synthetic datasets in Section 7 and show the importance of balancing application runtime and resultant accuracy when considering the potentially vast quantities of data that need to be processed. Within Section 8 we review our work and outline the future of our pipeline.

## 2 Related Work

Our pipeline can essentially be split into three principle components: feature extraction, clustering, and alignment, as illustrated in Figure 1. In terms of related works, the last two components have the most relevant supplemental work, so we focus on them.

### 2.1 Methods of Network Alignment

Within the field of alignment specifically, there are two primary approaches to merging networks, both depending on separate characteristics of the networks that are the target for alignment. These approaches can be generally defined as isomorphic alignment and attribute-based alignment.

Considering its popularity in research, isomorphic alignment tends to take the spotlight when it comes to application and uses. Specifically, isomorphic alignment revolves around the process of identifying matching nodes by exploiting global properties, i.e., graph structure, and adding attribute information when available, such as node weights. Some common approaches to representing commonality in networks are through spectral methods [10] or distributed belief propagation [1]. More recent research, however, is trending toward embedding structural representations of networks to use for alignment [8, 12, 13, 21].

While not mutually exclusive from the previously mentioned methods, another approach to aligning nodes is through the use of network attributes. In the context of graph representations this is often exhibited through direct use of node and edge attributes [17]. Alternatively, embedding a variety of discrete entity attributes can also be used for graph alignments [14].

Our work falls closer to pure attribute-based alignments, as the scope of our pipeline forces us to focus on simple representations of entity characteristics to find matches. Any kind of comprehensive structural or auxiliary enhancements to the entity attributes themselves could potentially increase the runtime of the overall pipeline to unreasonable lengths provided the massive amounts of data we seek to align in this paper. Therefore, we focus on correlating simple attributes in the most significant and efficient ways possible. And while the simplicity of our attribute-based approach is not novel, we introduce efficient, customizable, and scalable augmentations to the methods behind the alignment process itself that provide meaningful contributions to the field of entity alignment.

### 2.2 Clustering in Large-Scale Datasets

In addition to alignment, another important component of the pipeline is our clustering architecture, which is fundamental to handling alignments in big data. Like network alignment, clustering for large-scale data has seen extensive research given its usefulness in a myriad of fields. And although there is a large toolkit of varied approaches for clustering large amounts of data, we will focus on the two approaches most relevant to our paper: sampling-based clustering and parallel clustering.

Sampling-based clustering refers to the process of sampling some subset of the target data and using that subset for the clustering process itself. Within this process there are multiple methods of sampling that affect the quality of the clustering results. The standard sampling method for clustering tends to be uniform random sampling. One example can be found in [6] which shows how random sampling can improve the efficiency of their clustering processes. Another popular sampling method given the expanding ontology of big data is stratified sampling. In [18] stratified sampling is used to retrieve representative samples provided a large-scale dataset. These samples are partitioned into clusters using a fuzzy c-means algorithm. For completion, the remaining out-of-sample datapoints are then labeled given their similarity to partitioned clusters.

Parallel clustering, on the other hand, focuses on a split-apply-combine paradigm for large-scale data analysis. For clustering specifically, this involves scattering the dataset to a set of sub-tasks, clustering these partitions individually, and then consolidating the sub-task clusters into a single space. A key benefit of this approach is that each sub-task can be performed on a separate processor. The most popular representation of this methodology is MapReduce, a scalable framework that is particularly popular for the development of fast clustering algorithms [5, 11].

This paper builds off both the ideas of MapReduce and uniform random sampling to establish a unique take on large-scale clustering that specializes in scalability and customizability without sacrificing cluster quality.

## 3 Datasets

We leveraged multiple synthetic datasets to verify our pipeline. In each dataset, we refer to the separate activity graphs as "channels." The first two datasets were generated for the Modeling Adversarial Activity (MAA) DARPA Program [2]. We leverage two of the data channels in each dataset. The first dataset (DS1) contains about 80 million events, with 100 thousand unique cross-channel entities, while the second dataset (DS2) contains about 100 million events, with 5 million unique cross-channel entities. An event within a channel represents a passage of information from one unique entity to another. Each recorded event contains some set of information about the entities within the stream. For example: a source entity, a target entity, the channel that the transfer of information occurs in, and a unix timestamp. The channel pairs in each dataset can be aligned, with most entities in one channel having one (or more) matching counterparts within the opposite channel.

We also use a third data source, provided by VAST [20] for their 2019 challenge. This dataset contains call records, emails, purchases, and meeting records of a fictional company called Kasios International. Each of these transactional channels are represented by an event stream as well, following a format similar to the MAA datasets where each event provides a user source, user target, parent channel, and timestamp. There are 642,631 unique cross-channel users within the dataset, with an uneven number of them acting across the data source's mediums. In terms of scope, the largest of the channels is the email records, which contains 14.6 million events (or emails sent). This is followed by the call channel, which has 10.6 million events. The other two channels (company purchases and company meetings) contain 762,000 events and 127,000 events respectively. The range of time in which the activities of Kasios International occurs in is about 2.6 years.

These datasets represent significant challenges for alignment, as there is very limited data available; in particular, we are not leveraging entity names or similar attributes that are often used in such situations. Instead, we aim to correlate the temporal behaviors as a way to illustrate the scalable processing within our framework. Our focus is thus on scalability as opposed to absolute performance levels.

## 4 Feature Design for Alignment and Clustering

As indicated by the event attributes of both our data sources, each channel contains mostly temporal information: the timestamps at which a directional exchange of information occurs. Thus, our focus will be to primarily align cross-channel entities using correlated temporal activity. We consider two types of extracted temporal features: alignment-specific and cluster-specific. Alignment-specific features are the attributes that are used to correlate entities in the pipeline. Cluster-specific features, however, are used only during the clustering stage. These types of features will be more general measures, with lower dimensionality and lower discriminatory power than the alignment-specific features, allowing for more efficient cluster convergence.

### 4.1 Alignment-Specific Features

We define the set of $z$ available channels for a data source as $C = \{c_1, ..., c_z\}$, where $c_i$ represents the event stream. Given the myriad of options available for correlating two entities across a channel pair, a tempting solution may be to consider something along the lines of learning network embeddings in order to gain insight into latent structural characteristics. However, while this would most likely yield feature vectors that could be correlated across channels with high accuracy, the huge quantity of data that we must process prevents most arbitrary implementations of such solutions, especially if exceptionally powerful computers are not part of the available toolkit. Given these limitations, feature extraction methods with low computational complexity are the best to set the groundwork.

We find that while seemingly elementary, the idea of direct pairwise time series correlation is a reliable alignment paradigm for the data under consideration. One of the prime advantages is the feature extraction method itself, which counts the number of events in different time intervals, which we refer to as time bins. Given the event streams of $C$ and a specified time delta $\Delta t$ that defines the time bin range, the assembly of the temporal feature involves a simple counting procedure that enables us to process all event streams in a single pass, binning the activity for each entity by $\Delta t$. This extraction method results in a set of sparse features $\mathcal{X} = \{X_1, ..., X_N\}$, where $X_e$ is the resultant binned time series for some entity $e$ and where $N$ is the number of entities in all channels in $C$. We define a binned time series as a set of event counts $X_e =$ $[x_e^{t_0}, ..., x_e^{\bar{T}}]$ where $x_e^t \in \mathbb{N}$ (the set of positive integers, including 0), and the time index $t$ indicates the time bin, where $t_0$ is the time of the first event in any channel and $\bar{T} = T/\Delta t$ is the time of the last event in any channel. The resultant feature can often be large, depending on $\bar{T}$, but will also be sparse, allowing for methods that can quickly align the entities.

## 4.2 Cluster-Specific Features

We are also interested in extracting attributes that can be mapped into a reduced feature space. While reduced features are not optimal for accurate cross-channel entity matching, these features are valuable for clustering, which can significantly reduce the number of cross-channel comparisons made, thus improving both the processing and memory footprint of the pipeline. Since the primary goal of clustering is to group channel-specific entities by some attribute with the goal of finding each entity's counterpart from the opposite channel within the same group, the target attribute should maintain a balance between discriminatory power and the dimensions of the feature space. If the feature is too specific then the cluster convergence time will greatly increase, but with limited discriminatory power our clustering algorithm won't form useful groupings. Mapping general attributes to set-length feature vectors is a good way to ensure general behavioral indicators are captured without yielding significantly large feature spaces. The general attributes we used for our datasets are the average event count $AEC(e)$ (the average number of events per $\Delta t$ for each entity $e$), average edge delta $DEC(e)$ (the average of the differences between $\Delta t$ event counts for entity $e$), and initial event $IE(e)$ (timestamp of the first event for entity $e$). And so, we have a fixed feature space of $\mathbb{R}^3$, where we say our reduced clustering feature is represented as

$$RedF = <AEC', DEC', IE'> \quad (1)$$

Where $f'$ denotes the variables of $RedF$ have been scaled by dividing by the maximum value over all entities.

We also use a second clustering feature that is generated via an embedding algorithm, where we can define the size of the embedding prior to runtime. Given our focus on temporal attributes, our second clustering feature is a time series embedding generated from $\mathcal{X}$. Following the methodology of [7], we embed $\mathcal{X}$ in Euclidean space using the Laplacian Eigenmap technique. The Laplacian we use for eigen decomposition is computed by $L = D - W$ where $W$ is the similarity matrix, defined as

$$W_{ij} = \begin{cases} e^{-\text{DTW}(X_i, X_j)/\tau}, & i \neq j \\ 0, & \text{otherwise} \end{cases} \quad (2)$$

Where arbitrary hyperparameter $\tau > 0$. Dynamic Time Warp (DTW) [22] is used as the distance metric, characterizing the similarity matrix. For diagonal matrix $D$ we say $D_{ii} = \sum_{j=1}^{n} W_{ij}$. Given the eigenvalues $\lambda$ and eigenvectors $q$ from the eigen decomposition of the Laplacian we find the first $p$ nontrivial eigenvectors such that $U = (q_i, q_{i+1}, ..., q_p)$ where $(0 < \lambda_i \leq \lambda_{i+1} \leq \cdots \leq \lambda_p)$. Euclidean positions for each associated time series are found by $EmbF = U^T$. The dimensions for our embedded feature $EmbF$ are $N \times p$ where $p \ll \bar{T}$. Therefore, we have an embedded feature with an adjustable size, allowing for greater fine-tuning during the clustering stage of the pipeline.

## 5 Clustering with the Super-Point Framework

The fundamental goal of clustering within the pipeline is to group cross-channel entities together by behavior in order to reduce the number of alignment comparisons that need to be made, limiting the alignment process to entities within their respective clusters. However, for the size of datasets that we are considering in this paper, directly implementing classical techniques like K-Means will take extended times to converge, especially when considering how temporal features will define the clustering space. Our solution to maintain a reasonable runtime is to use a method based on the split-apply-combine strategy so commonly implemented by MapReduce clustering variants. This process, referred to as Super-Point, was designed to enable scalability, customizability to new features, and cluster convergence speed while avoiding subset approximations or dependent secondary out-of-sample labeling processes.

Given some $\mathbb{R}^N$ variable set $\mathcal{X} = \{X_1, ..., X_N\}$ where $X_i$ is a feature extracted using the methods described in Section 4, and where $N$ is the number of all entities in all channels, we define $k$ number of clusters and $w$ number of available sub-tasks that we wish the observables to be scattered to. Now using uniform, random sampling we distribute $\mathcal{X}$ into $\frac{N}{w}$ number of subsets, which are then evenly queued into the available sub-tasks. Within each sub-task, a general clustering algorithm groups the partition into $k$-number of clusters for each channel. Each cluster has its mean $\mu$ calculated, resulting in a set of $k$ $\mu$ values for each channel. Once all sub-tasks are completed in parallel and the queue of partitions is empty, we obtain $k \cdot \frac{N}{w}$ number of $\mu$ values for each channel. These super-points are single points that act as best-fit representations of their respective partition's general location in summed space. Next, we run another general clustering process on the super-points themselves, finding $k$ global centroids. Finally, all of $\mathcal{X}$ is labeled according to relative distance from the global centroids. This step is done in a single pass, generating finalized clustering results without having to iterate over the entirety of $\mathcal{X}$ multiple times. Figure 2 shows an outline of the framework described above.

This method allows us to tailor the specific clustering algorithm within the super-point process based on what would provide the best performance for a given dataset. While K-Means might potentially have the best cluster convergence time within the partitions given the simplicity of the process, there are alternatives that might yield more representative super-points. For example, consider the use of a Gaussian Mixture Model (GMM) in the initial partitioned clustering stage. If we fit a GMM where k is the number of mixture components, we can then generate random samples from the fitted distribution. We can then assign a weight to these samples using their associated probabilities. The set of samples along with their weights can be used as the super-points for the global clustering process. We would expect that this would add robustness to the process because we are using a larger set of super-points (not just the means of the clusters), properly weighted by the data-driven likelihoods.

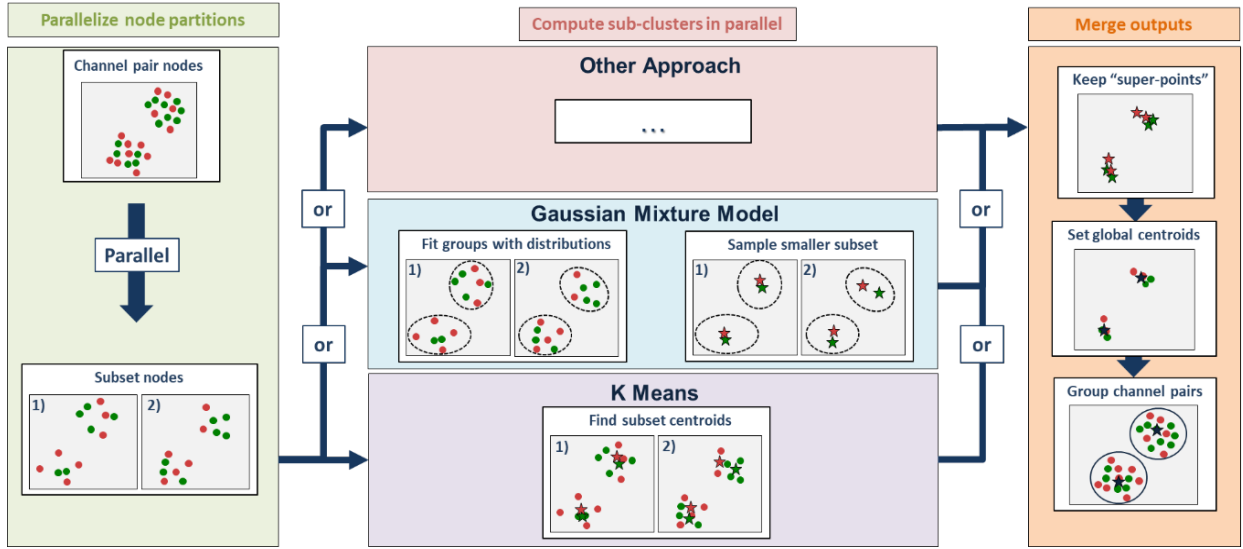

**Figure 2: Super-Point Clustering Framework**

While MapReduce has been used to implement many other efficient clustering algorithms, the super-point framework defines a process for incorporating a wide variety of clustering algorithms and distance metrics within the scope of fast, scalable computation that can be easily customized to best fit many different datasets.

## 6 Alignment and Alignment Likelihood Computations

The last step of our pipeline is to efficiently align our cross-channel entities by correlating features within a cluster. This stage can be split into two parts: the correlation of the cross-channel entities, and an optional alignment likelihood computation that can greatly improve the pipeline runtime.

### 6.1 Pairwise Alignment of Entities

Given corresponding clusters in two channels, we directly compare the feature vectors, $\mathcal{X} = \{X_1, \ldots, X_N\}$, keeping in mind that between, for example, channels 1 and 2, we know that an entity $u$ in one channel may or may not have one or more matching counterparts $v$ in the opposite channel. For a similarity metric, we use pairwise cosine similarity. When considering larger datasets, cosine similarity is often the best choice for feature comparison, as it is low complexity for sparse vectors, such as our temporal features. For two entities $u$ and $v$ we find $sim(u,v) = X_u \cdot X_v^T$ and apply an alignment threshold. Entities above the threshold are tagged as matches. Given the simplicity of the pairwise process, we can again utilize the MapReduce model, scattering subsets of $c_1$ and $c_2$ to w number of sub-tasks for alignment in smaller chunks. Within each sub-task, we can use the cluster labels to limit comparisons to the entities that are in corresponding clusters in each channel. Once each sub-task is complete, the resulting matching pairs for each process are consolidated into a single merged match, labeling which nodes are matches from channel 1 to channel 2. In order to ensure that our matches are robust, we run the same process again, but from channel 2 to channel 1. This process (referred to as the backward alignment) generates a separate match. To reconcile the two matches, we find the intersection between the two, outputting matches that have only occurred both for the forward alignment and the backward alignment. This intersection represents the final match.

### 6.2 Computing Alignment Likelihoods

For very large datasets, we may not be able to process all of the data in a single batch process. Given enough entities and events, the row and column lengths of the feature matrix that will be fed into the alignment process will increase, and the pipeline will inevitably either run out of memory. Our approach is to operate on temporal chunks of data and incrementally calculate alignment likelihoods.

To do this, the alignment likelihood computations output a likelihood score for each entity-entity pair indicating the likelihood that the pair in question is a match at the end of the alignment stage. Once the pipeline has completed a run for a segment of time, the alignments and likelihood scores are fed back into a new instance of the pipeline for the next time range that is queued. While running, this new instance updates the alignments and calculates new likelihood scores, which are then combined with the old likelihood scores from the previous run. At any point we can use the likelihoods to calculate a score for each entity-entity pair. Given some likelihood score threshold, the pairs with high enough scores are tagged as reliable matches. An illustration of this process is presented in Figure 3, where we partition the data into $N = T + 1$ day files and select sets of these files (multiple days) to process together. While alone these scores might be low (given the small range of $T$ each pipeline instance will receive) the result of combined scores across a set of time segments will capture growing trends of similarity between the appropriate entities. By allowing the pipeline to run on smaller segments of $T$, the overwhelming task of computing alignments for millions of entities with large temporal features are divided into smaller, more manageable runs where the temporal features are feasible to calculate and compare.

The alignment likelihood itself is computed by analyzing a small training set of computed features, split by channel. Our method is completely data driven—no labeling is needed, although we do assume that the training set includes true alignments.

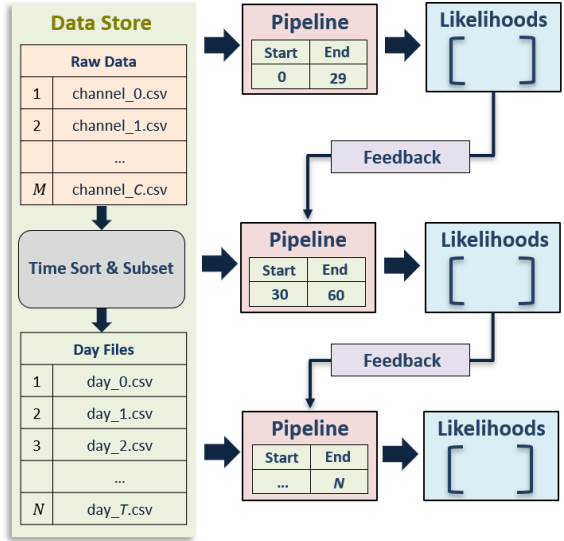

Figure 3: Flow of alignment likelihood scoring

For some channel pair 1 and 2 we can say $Y_1^{Train} \subset Y_1$ and $Y_2^{Train} \subset Y_2$. We calculate the cosine similarity between any two entities in opposite channels, then establish a hypothesis ($H_1$) that any two entities should be aligned and a null hypothesis ($H_0$) that any two opposite entities should not be aligned. We then calculate the probability of $H_1$ using the maximum similarity value per row of the training set.

$$P_1 = p(s|H_1), \quad (3)$$

$$s = \max_v(sim(u,v)), \quad (4)$$

$$u \in Y_1^{Train}, v \in Y_2^{Train} \quad (5)$$

Across $v$, this yields a distribution of $P_1$ (which reflects the hypothesis that entities are aligned) that can be fitted by a Rayleigh distribution. To compute the $H_0$ hypothesis, we generate a random, uniform sampling of all cosine similarities of some entity $v$, find the maximum of these scores, and compute the probability that the maximum preserves $H_0$.

$$P_0 = p(s|H_0), \quad (6)$$

$$s = \max_v(r_v(sim(u,v))) \quad (7)$$

Where $r_v(x_{uv})$ is an algorithm that randomly samples values of $x, x \in \mathbb{R}$ across a range of $v$. We use $r_v(x_{uv})$ in order to simulate a process of matching entity u against a set of opposite entities that will have similar temporal features, but no actual cross-channel matches. Like $P_1$, the distribution yielded by $P_0$ can also be fitted by a Rayleigh distribution. An example of the alignment likelihood distributions can be found in Figure 4.

As shown in Figure 4, the higher the similarity score, the more likely a pairwise alignment adheres to the $H_1$ hypothesis, giving distinct indicators as to where a likelihood score threshold should lie to maximize the correct identification of matches. In addition, the consideration of the overlap of the distributions can be used to characterize the quality of features used to generate the alignment values. A significant overlap of the two distributions can be used as an indicator of the weak discriminatory power of the involved features.

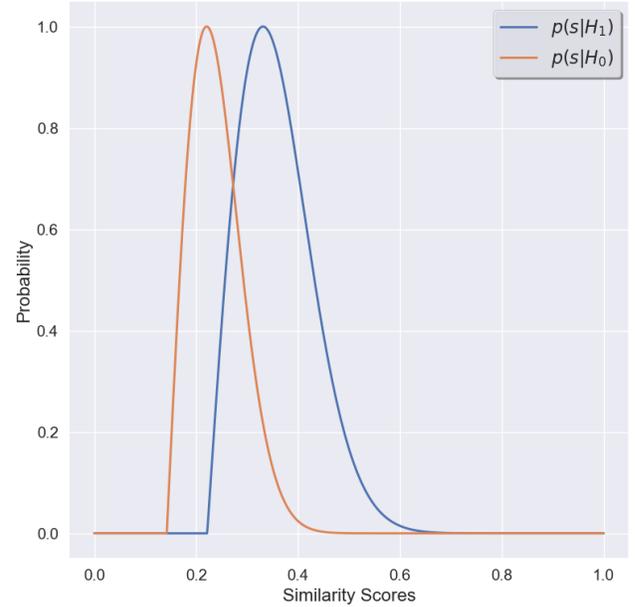

Figure 4: Distribution of Alignment Likelihoods

Given our final set of computed alignments from the previous section, we can use $P_1$ and $P_0$ to generate the likelihood scores for alignment, which we compute in log-space to ease consolidation of scores in the following pipeline runs. Outputting these log scores along with the alignment values upon the completion of a segment of $T$ can now inform the following instance of the pipeline as to the increasing or decreasing likelihood of a correct match for all entity-entity alignments.

## 7 Experiments and Results

To test the viability of our pipeline, we focus primarily on scalability (runtime), but also look at matching accuracy. In general, matching based on temporal information alone is extremely challenging.

## 7.1 Metrics

For our accuracy metrics, we have chosen to use metrics specifically designed for comprehensive evaluation of network alignments, as opposed to using set comparisons through Jaccard Similarity. These metrics, defined in [4], are referred to as Incorrectly Matched ($I_M$) and Incorrectly Not Matched ($I_{NM}$). The former is described as the percent of matches made that are not correct. The latter is described as the percent of matches that were missed by the alignment process.

## 7.2 Accuracy and Runtime Comparisons for VAST

To test the VAST dataset, we present results on the phone and purchases channels, which will be referred to as channels 1 and 3 from here on out, as these were the channels with the highest degree of temporal correlation. Using a machine of 72 Intel(R) Xeon(R) E5-2699 v3 CPUs at 2.30GHz with 388GBs of memory, we ran the VAST dataset through our pipeline. For testing, we consider three separate iterations of our pipeline: one run without clustering, one run with clustering using $RedF$ for the clustering features, and one run with clustering using $EmbF$ for the clustering features. For VAST we do not compute alignment likelihoods as total number of entities (~640K) is small enough to deemphasize the size of $T$ (~2.6 years), so there is no need to instantiate multiple pipeline runs.

Prior to runtime, we define the number of available sub-tasks to be $w = 22$. For clustering, we declare $k = 5$ and use the K-Means method for the partitioned clustering process within the Super-Point model. For the temporal feature extraction stage, we set $\Delta T = 3$ seconds due to the high degree of temporal correlation between our channels. The results for the three iterations of our pipeline, as evaluated by the metrics described in the previous section, are presented in Table 1. For notational simplicity, we present the metrics in terms of Matched Accuracy (MA) = $1 - I_M$, Forward Not Matched Accuracy (FNMA) = $1 - I_{NMF}$, and Backward Not Matched Accuracy (BNMA) = $1 - I_{NMB}$ on our plots.

As shown in Table 1, we have a consistent matching accuracy of 80%. In terms of alignment runtime, the pipeline keeps the process of computing the cosine similarities of a $640K \times 463K$ feature matrix and a $640K \times 27M$ feature matrix within a reasonable timeline of completion. In fact, clustering is able to reduce the alignment runtime somewhat, while losing essentially no accuracy.

**Table 1: Accuracy and runtime results for VAST**

| Pipeline | Channel Pair | MA | FNMA | BNMA | Runtime (s) |
|---|---|---|---|---|---|
| Without Clustering | (1,3) | 0.80 | 0.57 | 0.70 | 982 |
| Clustering (RedF) | (1,3) | 0.80 | 0.57 | 0.70 | 618 |
| Clustering (EmbF) | (1,3) | 0.80 | 0.57 | 0.70 | 464 |

## 7.3 Accuracy and Runtime Comparisons for DS1 and DS2

To process the MAA datasets, we segment the data into daily streams. Our pipeline is capable of ingesting one or more daily streams at once. On the DS1 dataset (100K entities), we ingest a month of data at a time, compute alignments on each ingested stream, and accumulate alignments over all data streams. We present the results in terms of likelihood scores against MA, FNMA, and BNMA, illustrated in Figure 5. As shown in the figure, the higher the likelihood score threshold, the better the accuracy of the pipeline. This makes sense, as the higher you set the likelihood score threshold, the more certain are the matches that are retained in the final match. With a likelihood score of 15, almost all matches are correct while correctly aligning ~65% of the entities that exist across the channels.

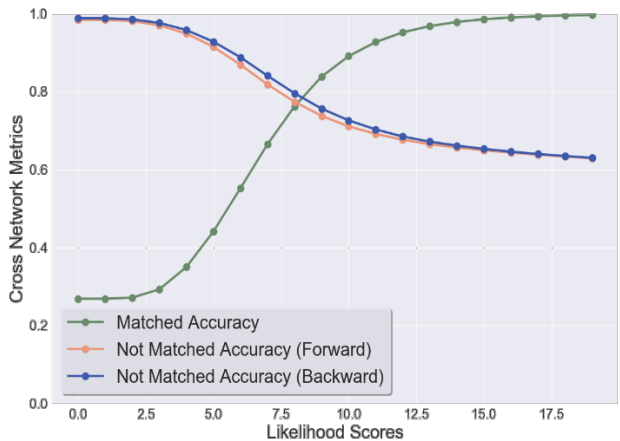

**Figure 5: Metrics as a function of likelihood scores on DS1**

In order to show runtime comparisons, we run on the DS2 dataset (5M entities), ingesting weekly data streams to reduce the overall memory footprint. We ran the pipeline through a range of progressively increasing subsets of the data to show the runtime as a function of the number of entities. The runtime results for this test are presented in Figure 6.

Figure 6 shows that runtimes are manageable up until 5 million entities per channel. At that point, the runtime increases to approximately 26 hours. Fortunately, utilizing clustering within the pipeline begins to make a significant difference around the 100K entity mark. Decreasing runtimes by a factor of 2, clustering brings the 5 million runtime mark from 26 hours down to feasible 10 hours of processing time. The clusters in this case are unbalanced (i.e., some clusters are much larger than others). With additional features we could achieve more balanced clusters, which would result in near ideal speedup in runtime.

The metrics are shown in Figure 7 for processing 9 weeks of the data, where if we threshold at a likelihood score of 1, we align ~640,000 entities correctly. The number of correctly aligned entities would increase as we incorporate more data into the system. Additionally, the clustering technique paired with the fairly high accuracy of the segmented likelihood approach demonstrates that the pipeline can scale to large datasets.

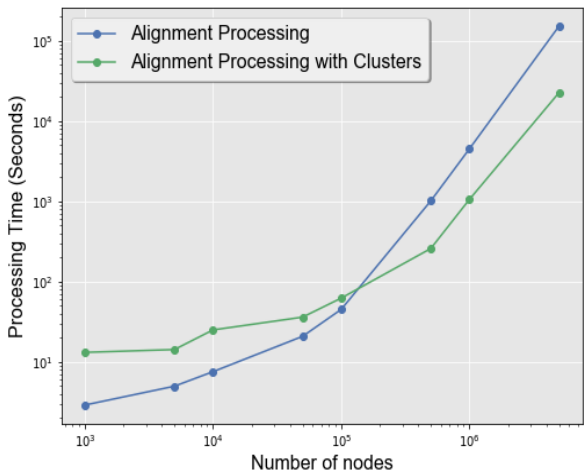

Figure 6: Alignment runtime comparisons for DS2

## 8 Conclusions and Future Work

In the world of big data, aligning massive amounts of entities is a prevalent and necessary problem to explore. To help address this issue, we introduced a set of methods, combined into a pipeline, that together enable aligning entities within large-scale networks with high accuracy.

First, we described a method of temporal feature extraction that was universal and easy to implement in the form of time binning. We paired this with two other feature extraction methods specifically for clustering, with the balance of reduced dimensionality and maintained discriminatory power in mind. Then we outlined a powerful clustering framework, built upon MapReduce that allowed for high customizability without losing any scalability. Using this framework, we can cluster temporal features by behavior, and find groups of potentially correlated entities, drastically reducing the number of pairwise computations that need to be made during alignment. Finally, we discussed methods for alignment. With the use of alignment likelihood scores, we showed that feeding economically sized chunks of the dataset incrementally into the pipeline results in appropriately accurate entity matches. The quality of results showed that together the methods outlined can lay a groundwork for future developments of similar alignment pipelines within the scope of big data.

In terms of future work, we will explore feature spaces on datasets with richer attributes and continue to fine-tune both the Super-Point framework and the alignment likelihood scores. Richer attribute spaces can provide a significant benefit to alignment accuracy when exploited and can help computation by creating more balanced clusters. The Super-Point clustering algorithm would likely be improved using more than one global update operation, whereby each iteration performs a new random partitioning of the data into subtasks. Additionally, we can compute sample statistics for each set of sub-task clusters using the mean and variance and perform weighted draws from those distributions for the global update step. Due to the nature of the likelihood scores, our pipeline is now capable of incorporating new data in a constant stream, continuously updating alignment hypotheses. We can also adapt our pipeline to read in streams of continuously updating data. With this setup we can test the limits of the current likelihood method and overall alignment process, and subsequently improve upon our established methods.

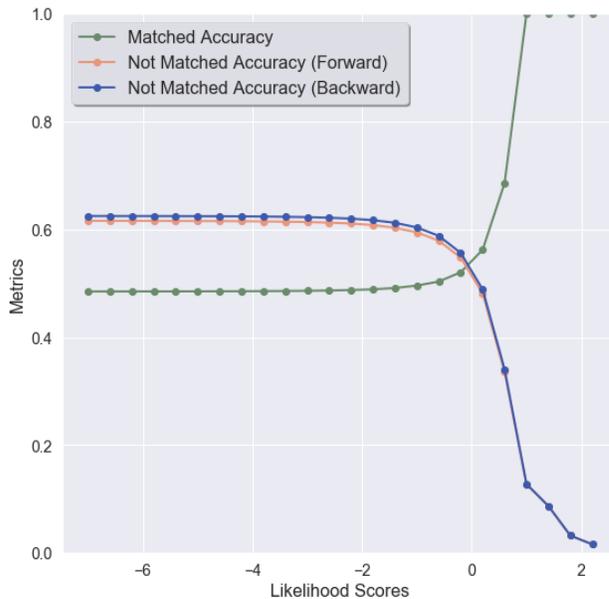

Figure 7: Metrics as a function of likelihood for DS2

## ACKNOWLEDGMENTS

This material is based upon work supported by the United States Air Force and DARPA under Contract No. FA8750-17-C-0156. The views, opinions and/or findings expressed are those of the author and should not be interpreted as representing the official views or policies of the Department of Defense or the U.S. Government.

Distribution A: Approved for Public Release, Distribution Unlimited